\documentstyle [11pt]{article}

\begin{document}

\begin{titlepage}

\bigskip

\bigskip

\bigskip

\bigskip

\bigskip

\title {\bf Cauchy Horizon Endpoints and Differentiability
\footnote{ \  1991 \   Mathematics
Subject Classification  53C50, 53C80, 83C75.} }

\author{\bf John K. Beem  and Andrzej Kr\'olak}

\maketitle

\end{titlepage}

\vspace*{.4in}

\centerline {\Large\bf  Cauchy Horizon Endpoints and Differentiability}

\bigskip

\centerline{John K. Beem$^{1}$ and Andrzej Kr\'olak$^{2}$
\footnote[3]{On leave of from Institute of Mathematics,
Polish Academy of Sciences, \'Sniadeckich 8, 00-950 Warsaw, Poland}} 

\bigskip

\noindent$^{1}$ Mathematics Department, University of Missouri-Columbia,
\\Columbia, MO 65211 USA. E-mail: mathjkb@showme.missouri.edu

\noindent$^{2}$ Max-Planck-Institute for Gravitational Physics,
Albert-Einstein-Institute, 
\\Schlaatzweg 1, 1447 Postsdam, Germany. 
E-mail:  krolak@aei-potsdam.mpg.de

\bigskip

\bigskip

\centerline  {\bf Abstract}

\smallskip

\begin{quotation}

\noindent
{\footnotesize   Cauchy horizons are shown to be differentiable at endpoints
      where only a single null generator leaves the horizon.
      A Cauchy horizon fails to have any
       null generator endpoints on a given
      open subset iff it is differentiable on the open subset and
      also iff  the horizon is (at least) of class $C^1$ on
      the open subset.  Given the null convergence condition, a
       compact horizon which is of class $C^2$ almost everywhere
      has no endpoints and is (at least) of class $C^1$ at all points.}
\end{quotation}
\bigskip

\centerline {\bf I. Introduction}

\bigskip

Cauchy horizons and black hole event horizons have been
extensively studied and used in relativity
[2 -- 6, 10 -- 14, 16, 17].
For general spacetimes, horizons may
fail to be stable under small metric perturbations,
however,
some sufficiency conditions for various stability questions
have been obtained  \cite{aaa}, \cite{chi}.  In the present paper, we will
consider
some differentiability questions for Cauchy horizons.

Let $(M, g)$ be a spacetime with a partial Cauchy surface $S$.  The
future Cauchy development $D^+(S)$ is the set of points of the
spacetime where, in theory, one may calculate everything in terms of
initial data on $S$.  The future Cauchy horizon $H^+(S)$ is the future
boundary of $S$.  We state our results in terms of the future horizon
$H^+(S)$, but similar results hold for any past Cauchy horizon $H^-(S)$.

Cauchy horizons are {\em achronal} (i.e., no
two points on the horizon may be
joined by a timelike curve) and this implies
that Cauchy horizons (locally)
satisfy a Lipschitz condition. This, in turn, implies that Cauchy horizons
are differentiable almost everywhere.  Because  they are differentiable
except for a set of (3 dimensional) measure zero, it seems that they
have often been assumed to be smooth except for a set which may be
more or less neglected.  However, one must
remember in the above that (1)
differentiable only refers to being differentiable at a single point and
(2)  sets of measure zero may be quite widely distributed.
In fact, Chru\'sciel and Galloway \cite{cag} have constructed
examples of  both a Cauchy horizon and a black hole event horizon where the
horizons fail to be differentiable on dense subsets.
Thus, they have constructed examples of horizons
which contain no open subsets on which they are differentiable at all points.
They point out that these examples raise definite questions concerning some
major arguments that have been given in the past where smoothness assumptions
have been implicitly assumed, compare \cite{cag}.
In the light of these new  examples, it is clear
that there is a real need for a deeper understanding of the
differentiability properties of horizons.

Each point $p$ of a Cauchy horizon $H^+(S)$ lies
on at least one null generator
\cite{hel}.   Since partial Cauchy surfaces are edgeless, a
null generator continues to lie on the horizon when it is extended in
the past.  However, null generators may or may not stay on the horizon
when they are extended in the future direction.  If a null generator leaves
the horizon, then there is a last point where it remains on the horizon.
This last point is said to be an {\em endpoint} of the horizon.

Endpoints where two or more null generators
leave the horizon are points where
the horizon must fail to be differentiable \cite{haa}, \cite{cag}.
In addition, Chru\'sciel and Galloway \cite{cag} have shown that Cauchy
horizons
are differentiable at points
which are not endpoints.  Furthermore, Chru\'sciel and Galloway pointed out
the need to resolve the differentiability issue for endpoints where
only one generator leaves the horizon.  In the present
paper, we show that Cauchy horizons are
differentiable at these endpoints.
This completes the classification of (pointwise) differentiability for
Cauchy horizons in terms of null generators and their endpoints.
However, this result raises the following question.  Is it always
true that the entire set of endpoints of a Cauchy horizon will have measure
zero?  We conjecture that  the answer should be affirmative based on
known examples.  Also, some support
for our conjecture is given by our proof that the set of endpoints with
only one generator is in the closure  of the set of endpoints with more
than one generator.

Restricting our attention to an open subset $W$ of
the Cauchy horizon $H^+(S)$ and assuming that the horizon
has no endpoints on the open set $W$, we find the horizon must
be differentiable at each point of $W$ and, in fact,
that the horizon must be at least of
class $C^1$ on $W$.  Conversely, we find differentiability on
an open set $W$ implies there are no endpoints on $W$.
In general, (pointwise) differentiability on an open set
yields class $C^1$, but not necessarily class $C^2$.
We give an example to
demonstrate that one may not conclude that either differentiability
or lack of endpoints on an open set $W$ imply class $C^2$ on $W$.
We also include a very simple example of a Cauchy horizon with
an endpoint where only one null generator leaves the horizon.

Hawking \cite{haa} has argued that under certain
conditions there should be no compactly generated Cauchy horizons.
These are horizons with a compact set $K$ such that all null generators
eventually enter and remain in this set $K$  when extended
in the past direction.  Hawking was interested in establishing arguments
against the possibility of having closed timelike curves in physical
spacetimes and, as he mentions, most of his
arguments apply to the special case of
compact horizons.  A key part of Hawking's paper involves flowing the
horizon back along null generators to get a contradiction.  We use his
notation and technique to find sufficient conditions for a compact
horizon to have no endpoints and to be at least of class $C^1$ at all points.
More precisely, we show that if one has the null convergence condition and a
compact horizon which is of class $C^2$ on an open set $G$ with complement
of measure zero, then the horizon has no endpoints and is (at least) of class
$C^1$ at all points.

\bigskip
\bigskip

\centerline   {\bf 2. Preliminaries}

\bigskip
								
Let $(M,g)$ be a spacetime. Although our results
hold for n-dimensional
spacetimes, we will only give the proofs in
the four dimensional case since
similar proofs hold in other dimensions.
Thus, we will take $M$ to be a smooth,
connected, four dimensional, Hausdorff manifold with a
Lorentzian metric, a countable basis, and a time orientation.
The Lorentzian metric $g$ will
have signature $(-, +, +, +)$.   	A {\em partial Cauchy surface}
$S$ will be a connected,
acausal, edgeless three dimensional submanifold of
$(M,g)$,  compare \cite{hel},  \cite{haa}, \cite{aac}.
The {\em future Cauchy
development} $D^{+}(S)$ consists of all points $ p \in M$ such that each past
endless and past directed causal curve from $p$ intersects the set $S$.
The {\em future Cauchy horizon} is
$H^{+}(S) = \overline{(D^{+} (S))} -I^{-}(D^{+}(S))$.
Let $p$ be a point of the Cauchy horizon.  It is well known, that there
is at least one null generator of $H^{+}(S)$ containing $p$.  Each null
generator is at least part of a null geodesic of M.   When a null
generator of $H^{+}(S)$ is extended into the past, it continues to lie on
the horizon, compare [12, p. 203].   However, if a null generator is
extended into the future it may have a last point on the horizon
which then said to be an {\em endpoint} the
horizon.  We will define the {\em multiplicity} of a point $p$
in $H^{+}(S)$ to be
the number of null generators containing $p$.  Points of the horizon
which are not endpoints must have multiplicity one.  The multiplicity
of an endpoint may be any positive integer or infinite.
We will call the  set of endpoints of multiplicity
two or higher the {\em crease set}, compare  \cite{cag}.
	
Consider any fixed point $p$ of the Cauchy horizon $H^{+}(S)$ and let
$x^{0},x^{1}, x^{2},x^{3}$ be local coordinates defined on an open set
about $p = (p^0, p^1, p^2, p^3)$. Let $H^+(S)$ be given near
$p$ by an equation of the form

$$
x^0 = f_H(x^1, x^2, x^3)
$$
\smallskip

\noindent
The horizon $H^+(S)$ is {\em differentiable} at the point $p$ iff
the function $f_H$ is differentiable at the point $(p^1, p^2, p^3)$ using
the advanced calculus definition of differentiability [18, p. 212].
In particular, if $p = (0, 0, 0, 0)$ corresponds to the origin in the given
local coordinates and if

$$
\Delta x = (x^1, x^2, x^3)
$$
 \smallskip

\noindent
represents a small displacement from $p$ in the $x^0 = 0$ plane, then
$H^+(S)$
is differentiable at $p$ iff one has

$$
f_{H}(\Delta x) = f_{H}(0) + \sum a_{i}x^{i} + R_H(\Delta x)
 = 0 + \sum a_{i}x^{i} + R_H(\Delta x)
$$
\smallskip

\noindent
where the ratio $R_H(\Delta x)/| \Delta x |$
converges to zero as $| \Delta x |$ goes to zero.
Here we use

$$
|\Delta x| = \sqrt{ (x^{1})^{2} + (x^{2})^{2} + (x^{3})^{2} }.
$$
\smallskip

\noindent
Note that even though
$| \Delta x |$ represents a coordinate dependent length, the Cauchy surface
$H^+(S)$ will be differentiable or not differentiable at $p$ independent of
the particular choice of local coordinates.
Of course, when $H^+(S)$ is differentiable at $p$, one has
$a_{i} = \partial f_{H} / \partial x^{i}$ evaluated
at the origin for   $i  = 1,2, 3$.  The surface $H^+(S)$ is of class
$C^r$ on an open neighborhood of $p$ iff $f_H$ is of class $C^r$ on an open
neighborhood
of the origin.

If $H^{+}(S)$ is differentiable at the point
$p$, then there is a well defined 3-dimensional linear subspace  $N_0$
in the tangent space $T_{p}(M)$ such that $N_{0}$ is tangent to the
3-dimensional surface
$H^{+}(S)$ at $p$.
In the above notation a basis for $N_{0}$ is given by
$\{a_{i} \partial / \partial x^{0} + \partial / \partial x^{i} \ | \  i =
1,2, 3 \}$.
It is clear that if local coordinates $x^0, x^1, x^2, x^3$ have been chosen
as above,
then the partials $a_i = \partial f_H / \partial x^i$ at $p$ are determined
by $N_0$
assuming, of course, that $H^+(S)$ is differentiable
at $p$ and is given by $x^0 = f_H(\Delta x)$  near $p$.
Notice that the given tangent plane $N_0$ cannot be spacelike since
there is at least one null generator containing $p$.  Also, the tangent
plane $N_0$ cannot be timelike since otherwise $p$ would be in
chronological past of some points of $H^{+}(S)$.   Thus, $N_0$ is null (i.e.,
of signature $(0, +, +)$) and it follows that there is a one
dimensional linear subspace $L_0$ in $N_0$ such that $L_0$ is null.
In fact, $N_0 = L_0{}^{\perp}$
is the orthogonal space to $L_0$ and is thus uniquely determined by $L_0$.
Given a tangent vector $X$ at $p$ with
$X \in N_0$, then $X$ belongs to $L_0$ iff  $g(X, Y) = 0$ for all $Y \in
N_0$.

At a point
$p$ where $H^{+}(S)$ is differentiable there can only be one  null generator
though $p$ since two different null generators though $p$ would yield a
two dimensional timelike plane
lying in the (3-dimensional) tangent plane to the horizon and this would
imply the
existence of a timelike plane tangent to $H^{+}(S)$ at $p$ in contradiction to
the above.   Thus at points with
two or more null generators of $H^{+}(S)$, the horizon $H^{+}(S)$ cannot be
differentiable, compare \cite{cag}.
In fact, if  $p$ is a point of the horizon having two or
more null generators, then all null generators through $p$ must leave the
horizon at p when traversed in the future direction.  This follows
since if $\gamma_{1}(t)$ and $\gamma_{2}(t)$ are future directed null
generators with $\gamma_{1}(0) = \gamma_{2}(0) = p$ and if
$\gamma_{1}(t)$ remains on $H^{+}(S)$ for some $\gamma_{1}(t_{1})$ with
$t_{1} > 0$,
one has that all points of $\gamma_{2}(t)$ with $t < 0$ are in the chronological
past of $\gamma_{1}(t_{1})$ since one can traverse $\gamma_{1}(t)$ backward
to $p$,
 make a corner at $p$ to head in the past direction of the
null geodesic $\gamma_{2}(t)$ and then
traverse this null geodesic backward.
This would imply the points $\gamma_{2}(t)$ for $t < 0$ are in the set
$D^{+}(S)$
and not on the horizon, in contradiction.

We now give an example of an endpoint $p$ of
multiplicity one.

\bigskip
\bigskip

\noindent {\bf Example 2.1.}   Let $M_{0} = L^{3}$ be
three dimensional Minkowski spacetime
with coordinates $(t, x, y)$ and $g = -dt^2 + dx^2 + dy^2$.  Remove the
parabola $y = x^2$ from the $t = 0$ plane to obtain the spacetime $M$.  Let
the partial Cauchy surface $S$ be the portion of the x-y plane which is
interior to the parabola $y = x^2$ (i.e., $S = \{ (0, x, y)  \  |  \  y >
x^2 \}$).
The curvature $k$ of the curve $y = x^2$ has a maximum value of $2$ at $x = 0$
and is always less than 2 for points on the parabola other than the
origin.  The osculating circle at the origin to the curve $y =
x^2$ is a circle in the x-y plane of radius $1/k = 1/2$ and center at $y =
1/2$ on the y-axis.  The chronological past of the point $(1/2, 0, 1/2)$
intersected with the x-y plane is the interior of this osculating
circle and lies in the set $S$.  It follows that the null geodesic $x = 0,
y = t$ contains a null generator of $H^{+}(S)$.  In particular, the curve
$c(t) = (t, 0, t)$ for  $0 < t < \infty$ is a null geodesic which lies
on the horizon for $0 < t \leq 1/2$.
Notice that  if one intersects the x-y plane with the
chronological past of $(t, 0, t)$ for $t > 1/2$, then one obtains some
points outside of the above partial Cauchy surface $S$.  Thus, the null
generator $c(t) = (t, 0 , t)$ must leave the horizon at $t = 1/2$ and one finds
that $(1/2, 0, 1/2)$ is an endpoint of the horizon.
However, the endpoint $p = (1/2, 0, 1/2)$ has no other null generators since
all other null geodesics through this point intersect the set $S$.  Thus,
$p$ is an endpoint of multiplicity one as desired.  In this example one
may directly check that $H^{+}(S)$ is differentiable at the point $p$.  The
crease set is a curve which lies above the points on the y-axis for which
$ 1/2 < y < \infty$.

\bigskip
\bigskip

In the next example we obtain a horizon which has an
open set $W$ where the horizon is of class $C^1$, but where there are some
points in this open set $W$ where the horizon fails to be of class $C^2$.  In
particular, there is a certain null generator and the horizon fails to
be of class $C^2$ along this null generator.

\bigskip

\bigskip

\noindent {\bf Example 2.2.}    Let $M_{0} = L^{3}$.  The  spacetime
$M$ will be obtained by removing two half lines and one fourth of a circle. From
$M_0$ remove the set in the $t = 0$ plane given by  $ y = -1$ and $0 \leq
x$. Then
remove the part of the circle $ x^2 + y^2 = 1$ for which $ x < 0$ and $y < 0$.
Finally, remove the half line in the $t = 0$ plane given by $x = -1$ and $0
\leq y$.
The partial Cauchy surface $S$ will be the open set in the x-y plane
corresponding to
the interior of the convex hull of the deleted set.
The null generators corresponding to the quarter circle focus at the point
$(1, 0, 0)$.  The null generators corresponding to points on the line $y =
-1$ for
sufficiently small values of $t$ lie on lines of the form $c(t) = (t, x, t
- 1)$
where $0 \leq x$.  It follows that for values of $x$ and $y$ near $x = 0$ and
$y = -1/2$ the height $t$ of the Cauchy horizon is given by $t = f_H(x, y)$
where
$f_H(x, y) = 1 + y$ for $0 \leq x$ and also $f_H(x, y) = 1 - (x^2 +
y^2)^{1/2}$ for
$x < 0$. It is easily checked that $f_H$ is of class $C^1$ near $(0, - 1/2)$, but that
$f_H$ fails to be of class $C^2$ along the y-axis near $(0, - 1/2)$.  It
follows that
there is an open set $W$ in $H^+(S)$ about the point $(1/2, 0, - 1/2)$ where the
horizon is of class $C^1$, but not of class $C^2$.  In this example there are no
endpoints of multiplicity one.  The crease set consists of the closed half line
$t = y + 1 = x + 1$ for $0 \leq y < \infty$ and is a spacelike curve.

\bigskip
\bigskip

Both  Examples 2.1 and 2.2 may be changed to
four dimensional examples using a cartesian
product with a positive definite $R^1$ corresponding to
the z-axis.  In particular, to modify
Example 2.1, one may take $M_0 = L^4$, delete the
two dimensional set $\{ (0, x, y, z)  \  |  \   y = x^2 \}$,
and use the three dimensional partial Cauchy surface
$S = \{ (0, x, y, z)  \  |   \  y > x^2 \}$. This generates a
horizon $H^+(S)$ with an endpoint $p = (1/2, 0, 1/2, 0)$ of
multiplicity one.  To modify Example 2.2, one
may take $M_0 = L^4$ and
delete the set formed by
the cartesian product of the originally deleted set with the z-axis.
Of course, in this four dimensional example, one takes the new
partial Cauchy surface to be the cartesian product
of the original Cauchy surface with the z-axis.

\bigskip
\bigskip
\bigskip

\centerline {\bf 3. Differentiability of Cauchy Horizons}

\bigskip
\bigskip

One may always represent tangent null directions using a normalization
based on an auxiliary positive definite metric $g_{pos}$.
Given a fixed point $p$ on a null
geodesic  $\gamma$, chose the uniquely defined tangent vector $V$ to
$\gamma$ at $p$ which is future pointing and satisfies $g_{pos}(V, V) = 1$.
With this normalization, it is clear that the null directions at
$p$ form a compact set homeomorphic to $S^2$.
Of course, we are identifying the null direction given by a
null vector $V$ as the same as that given by
$\alpha V$ for nonzero values of
$\alpha $.  Thus, in particular, the null direction represented by
$V$ is the same as the null direction represented by $- V$.
Note that if one is given any compact subset $K$ of $M$, the
set of null directions attached at points of $K$ forms a compact
set.

Let $\gamma_n$ be a sequence of null generators of $H^+(S)$ and let
$p_n \in \gamma_n$ for each n.  Assume the sequence
$\{ p_n \}$ converges to $p$ and let $\gamma$ be a null
generator at $p$.  We will say the sequence $\gamma_n$
{\rm converges} to $\gamma$ if there is a sequence of
null vectors $\{ V_n \}$ converging to the null vector
$V$ such that each $V_n$ is tangent to the corresponding
generator $\gamma_n$ at $p_n$ and $V$ is tangent to $\gamma$ at $p$.
Notice that in this definition we do not assume that the points
$\{ p_n \}$ have unique null generators and we do not assume
that the point $p$ has a unique null generator.
In particular, we allow for
the possibility that some (or all) of $\{ p_n \}$ and $p$ may be
endpoints.

We now show that if $p$ is a point of multiplicity one, then any
point on the horizon sufficiently
close to $p$ must have all of its null generators
close to the unique null generator containing $p$.

\bigskip
\bigskip

\noindent {\bf Lemma 3.1.}
{\em Let $p$ be a point of multiplicity one on the horizon $H^+(S)$.
If $\{ p_n \}$ is a sequence of points on the horizon
converging to $p$ and if for each $n$ a null generator $\gamma_n$
containing the point  $p_n$  has been chosen, then the sequence
$\gamma_n$ converges to the null generator $\gamma$ containing $p$.}

\smallskip

\noindent {\em Proof.}
Assume by way of contradiction that the sequence $\{ p_n \}$
converges to $p$, but that the  null directions defined by
the generators $\gamma_n$ do not converge to the null direction
defined by $\gamma$.  Using compactness, it follows that
there must be subsequence $\{ j \}$ of the sequence $\{ n \}$  with the
$\gamma_j$  directions converging to
a null direction at $p$ different from that defined by $\gamma$.  This
yields a second null generator at $p$, in contradiction to the assumption
that $p$ was of multiplicity one.  $\Box$

\bigskip
\bigskip

Clearly, Lemma 3.1 implies that if $W$ is an open subset of
the horizon and if each point of $W$ has multiplicity one, then the
null generators move in a continuous fashion on $W$.
More precisely, we have the following result.

\bigskip
\bigskip

\noindent {\bf Lemma 3.2.}
{\em Let $W$ be an open subset of the future Cauchy horizon $H^+(S)$.
If each point of $W$ has multiplicity one, then
there is a (nonvanishing) null vector field $V$ tangent to
null generators and defined on $W$ such that $V$ is continuous.
Furthermore, the vector field $V$ may be taken to be future
pointing.}

\bigskip
\bigskip

We are now able to show that for open subsets of the
horizon, differentiability implies
(at least) class $C^1$.  Example 2.2 shows that
differentiability on an open set does not, in general,
imply class $C^2$ and hence
the following is the best possible result one may obtain without additional
assumptions.  Recall that if a surface is of class $C^r$ for $r > 1$, then
it is also of class $C^1$.

\bigskip
\bigskip

\noindent {\bf Proposition 3.3.}
{\em If $W$ is an open subset of $H^+(S)$ and if $H^+(S)$ is differentiable
at all points of $W$, then $H^+(S)$ is of class $C^r$ on $W$ for
some $r \geq 1$.}

\smallskip

\noindent {\em Proof.}
A point where the horizon is differentiable must be a point of multiplicity
one.
Hence Lemma 3.2 yields the existence of a  null vector field
$V$ on $W$ which is continuous, tangent to $H^+(S)$
and serves to define the generator directions on $W$.
Let $p \in W$ and, as in
Section 2, introduce local coordinates $x^0, x^1, x^2, x^3$ near $p$
with $\partial / \partial x^0$ future timelike.
Let $H^+(S)$ be given locally
by $x^0 = f_H(\Delta x)$.  The partial derivatives of $f_H$ exist
near $p$  since $H^+(S)$ is differentiable near $p$.  Notice that
the null vector field $V$ determines the orthogonal space $V^{\perp}$
which is the null tangent plane  to
$H^+(S)$ at each point near $p$.  Furthermore, the tangent plane
$V^{\perp}$ determines the first
order partial derivatives
of $f_H$.  The continuity of $V$ yields the continuity
of $V^{\perp}$ and hence the continuity of the first order
partial derivatives of $f_H$.  Thus,
$f_H$ is at least of class $C^1$ which is
equivalent to $H^+(S)$ being at least of class $C^1$,
as desired.  $\Box$

\bigskip
\bigskip

Fix a point $p$ on $H^+(S)$ of multiplicity one.  Choose normal
coordinates $x^0, x^1, x^2, x^3$ with $p$ corresponding to the origin,
with $\partial / \partial x^0$ future timelike at $p$, and with the
future direction of the unique null generator at $p$ corresponding to
the direction of $\partial / \partial x^0  + \partial / \partial x^1$.
Assume also that the natural basis is orthonormal at $p$. As before,
let $H^+(S)$ be given near $p$
by $x^0 = f_H(\Delta x)$.  If $H^+(S)$ is
differentiable at $p$ one has $\partial f_H / \partial x^1 = 1$ and
$\partial f_H / \partial x^2 = \partial f_H / \partial x^3 = 0$ at $p$.
Assuming that $f_H$ is of class $C^r$ for $r \geq 1$  near $p$, define
new coordinates $z^0, z^1, z^2, z^3$ by $z^0 = x^0 - f_H(\Delta x)$,
and $z^i = x^i$ for $i = 1, 2, 3$.  Locally $H^+(S)$ corresponds to the
set $z^0 = 0$.  Note that $z^0, z^1, z^2, z^3$ are local coordinates
of class $C^r$ in terms of the original $x$ coordinates.

\bigskip
\bigskip

\noindent {\bf Proposition 3.4.}
{\em If a  Cauchy horizon $H^+(S)$ is  differentiable on an  open subset $W$,
then the horizon has no endpoints on $W$.}

\smallskip

\noindent {\em Proof.}
The horizon must be (at least) of class $C^1$ on $W$ by Proposition 3.3.
Fixing $p \in W$, we may use the above
local coordinates $z^0, z^1, z^2, z^3$ which are
at least of class $C^1$ and the horizon is given near $p$ by $z^0 = 0$.
>From Lemma 3.2 it follows that there
is a continuous future directed null
vector field  $V$ tangent to $H^+(S)$ near $p$.  Then
$V = \sum y^i(z^1, z^2, z^3) \partial / \partial z^i$  where
$y^i = y^i(z^1, z^2, z^3)$ are continuous functions for $i = 1, 2, 3$.
Consider the system

$$
dz^i / dt = y^i(z^1, z^2, z^3)
$$
for $i = 1, 2, 3$.
This is a system which we may regard as being defined on
the coordinate plane $z^0 = 0$ (i.e., on
the horizon) and solving the system
corresponds to finding integral curves of the vector field $V$.
However, the above system does not necessarily satisfy a Lipschitz
condition and thus, in
principle, might fail to have unique solutions.  Nevertheless, because of
the continuity of the functions $y^i$, one may apply the Cauchy-Peano
Existence Theorem [8, p. 6] to obtain a class $C^1$
solution $c(t)$ with $c(0) = p$.  Notice that $c(t)$ is a null curve
always tangent to $H^+(S)$ because $c^{\prime}(t)$ is always equal to
the vector field $V$ at the corresponding point $c(t)$.
Moving along the curve $c$ a little to the past, one obtains a point
$r = c(t_1) \in H^+(S)$ with $t_1 < 0$. Similarly, moving along $c$ a little
to the future one obtains $q = c(t_2) \in H^+(S)$ with $t_2 > 0$.
Traversing the curve $c$
backward from $q$ to $r$ one finds it must
be a null geodesic lying on the horizon
since if it were not, then $r$ would be in
the chronological past of $q$ and
this would yield  a contradiction to the
achronality of $ H^+(S)$.  Since there
is a unique null direction tangent to $H^+(S)$ at $p$, the
curve $c$ (at least) restricted
to the domain $[ t_1, t_2 ]$ must lie on the generator containing $p$ and it
follows that $p$ is not an endpoint, as desired.   $\Box$

\bigskip
\bigskip

We note in passing that the proof of Proposition 3.4 shows
that, in fact, the system $dz^i / dt = y^i(z^1, z^2, z^3)$ has unique
solutions given an initial point $p \in W$ and that each solution $c(t)$
lies on  (part of) a null generator.

Fix a point $p$ of multiplicity one on $H^+(S)$ with null direction $L_0$
tangent to the unique null generator at $p$.  Let $Y_0, Y_1, Y_2, Y_3$ be
an orthonormal
basis at $p$ with $Y_0$ future pointing timelike and with $Y_0 + Y_1$
future pointing
null in the direction of $L_0$.  Then $Y_2$ and $Y_3$ lie in the orthogonal
space to
$L_0$.  Let $X_0, X_1, X_2, X_3$ be a new basis of $T_pM$ given by
$X_0 = Y_0$, $X_1 = Y_0 + Y_1$, $X_2 = Y_2$, and $X_3 = Y_3$.   Take normal
coordinates $x^0, x^1, x^2, x^3$
centered at $p$ determined by the basis $X_0, X_1, X_2, X_3$.  Then
$\partial / \partial x^0$ is timelike future pointing,
$\partial / \partial x^1$ future pointing in
the null direction $L_0$ and both
$\partial / \partial x^2$ and $\partial / \partial x^3$
are in the orthogonal space to
$L_0$.  Of course, these normal coordinates fail to
be orthonormal at $p$ since, in particular, $\partial / \partial x^1$ is
null.  Notice that the metric tensor $g$ has components (in the $x$
coordinates) which at
$p$ have all zero values except for $g_{00}(p) = g_{01}(p) = g_{10}(p) = -
1$ and
$g_{22}(p) = g_{33}(p) = 1$.   The  $x^1$ axis is a geodesic which corresponds
to a null generator of $H^+(S)$ and (at least) the origin and
negative $x^1$ axis lie on the horizon.  The point $p$ is an endpoint
iff points on the positive $x^1$ axis do not lie on the horizon.  In the
following we allow $p$ to be either an endpoint or a nonendpoint.
Fix a point $r = (0, t_1, 0, 0)$ on the negative $x^1$ axis
and lying in a convex
normal neighborhood of $p$.  Thus, $t_1 < 0$.  Since we have used normal
coordinates, we have $p = exp_r(V_0)$ where
$V_0 = | t_1 | \partial / \partial x^1$
and $V_0$ is a null vector attached at $r$.  If one considers
the exponential map at $r$ restricted
to all null vectors at $r$ in some small
neighborhood of $V_0$, then one obtains
a smooth null surface which passes through $p = (0, 0, 0, 0)$.
Since $\partial / \partial x^1$ at the origin is a null
vector tangent to this null surface, it follows that
the  tangent plane to this null surface at the point $p$ is
the orthogonal plane to $\partial / \partial x^1$ which
is also the tangent plane to the
coordinate plane $x^0 = 0$ at $p = (0, 0, 0, 0)$.
Let $N(V_0)$ be a sufficiently small neighborhood of $V_0$ in
the tangent bundle $TM$ and
let $U(r)$ be a (small) 3 dimensional neighborhood
of $r$ in the coordinate plane $x^1 = t_1$.
Then points of $U(r)$ are of the
form $u = (u^0, t_1, u^2, u^3)$ where the values of
of $u^0, u^2$ and $u^3$ are all close to zero.
For each fixed $u \in U(r)$ one
generates a smooth null surface $S(u^0, u^2, u^3)$
defined near the origin by using the exponential map at $u$
and restricting
$exp_u$ to  null vectors in the set $N(V_0)$ attached at $u$.
An element  $V \in N(V_0)$ attached at $u \in U(r)$ may be  represented
in local coordinates as
$V = V^0 \partial / \partial x^0 + \sum V^i \partial / \partial x^i$.
Thus, using g(V, V) = 0 to determine the
$V^0$ component in terms of the $V^1, V^2, V^3$ components, one
may parametrize our six dimensional domain space
by $u^0, u^2, u^3, V^1, V^2, V^3$.  Here the values of
$u^0, u^2, u^3, V^2$, and  $V^3$ are close to zero and the
value of $V^1$ is close to $|t_1|$.
Note that tangent vectors to this
six dimensional space may be expressed in terms of a basis
of the form $\partial / \partial u^0, \partial / \partial u^2,
\partial / \partial u^3, \partial / \partial V^1, \partial / \partial V^2,
\partial / \partial V^3$.   Consider the map

$$
E(u, V) = exp_u(V)
$$

\noindent
taking elements of the domain space
to points of $M$ near $p$. This map
may be written as
$E(u, V) = (E^0, E^1, E^2, E^3)$ where each component
$E^{\mu}(u^0, u^2, u^3, V^1, V^2, V^3)$  is a real valued function of
six variables.   The reader will note a slight abuse of
notation in the above.  For example, we
will use $u$ to  denote both
the point $(u^0, t_1, u^2, u^3)$ in $M$ and
also the corresponding coordinates $u^0, u^2, u^3$ in
our domain space.  A similar comment holds for our use of $V$.
Notice that if $u$ is chosen to be the point
$r$, then the map $E(r, -)$ takes the null vectors attached at $r$ to
the null surface $S(0, 0, 0)$ which we noted above is
tangent to the coordinate plane $x^0 =0$ at
$p$.  In our domain space the vector
$V_0$ in $T_rM$ corresponds to
$(0, 0, 0, | t_1 |, 0, 0)$.  Notice that the derivative of
the exponential map for $u = r$ (i.e., $(exp_r)_*$ )
is nondegenerate at $V_0$
since $r$ was chosen close to $p$.    Furthermore,  $(exp_r)_*$
takes the null hyperplane at $V_0$ tangent to the null vectors at $r$
to the null plane tangent to $x^0 = 0$ at $p = (0, 0, 0, 0)$.
Thus, for fixed $u = r$ and $V = V_0$, $E_*$ takes
the three dimensional space with basis
$\partial / \partial V^1, \partial / \partial V^2, \partial / \partial V^2$
corresponding  to certain vectors tangent to our six dimensional domain
space and attached to this domain space at
$V_0 = (0, 0, 0, | t_1 |, 0, 0)$
in a nonsingular fashion to the
three dimensional space with basis $\partial / \partial x^1,
\partial / \partial x^2, \partial / \partial x^3$ of vectors
tangent to $M$ and attached at
$p$.  Thus, using $x^i = E^i(u^0, u^2, u^3, V^1, V^2, V^3)$ one finds
that the three by three
matrix [ $ \partial x^i / \partial V^j $ ] is
nonsingular when evaluated at the point
$(0, 0, 0, | t_1 |, 0, 0)$ of our domain space.
Using the Implicit Function Theorem \cite{lan}, we may thus
solve for the $V^1, V^2, V^3$ variables in terms of the
$u^0, u^2, u^3, x^1, x^2, x^3$ variables.  Hence, one
obtains three $C^{\infty}$ functions
$V^i = V^i(u^0, u^2, u^3, x^1, x^2, x^3)= V^i(u, \Delta x)$.
We will let $F$ be the $x^0$ component of $E$ using the
$u^0,u^2, u^3, x^1, x^2, x^3$ variables.  In other words, we
define $F$ to be the real valued function

$$
F(u^0, u^2, u^3, x^1, x^2, x^3) =
E^0(u, V^1(u, \Delta x), V^2(u, \Delta x), V^3(u, \Delta x)).
$$

\noindent
Then, for fixed $u$, the surface $S(u^0, u^2, u^3)$
is given near the origin by
$x^0 = F_u(x^1, x^2, x^3)$ where
$F_u(\Delta x) = F(u^0, u^2, u^3, x^1, x^2, x^3)$ is a
smooth function of six variables.
Still holding $u \in U(r)$ fixed, select a point
$q = (q^0, q^1, q^2, q^3)$ near the origin on the
surface $S(u^0, u^2, u^3)$.  Then
one has $q^0 = F_u(q^1, q^2, q^3) = F_u(\Delta q)$. Now expand $F_u$
about this point (holding the $u$ coordinates fixed) to obtain

$$
F_u(x^1, x^2, x^3) = q^0 + \sum (\partial F_u / \partial x^i ) (x^i - q^i)
               + R(u, q,  \Delta (x - q))
$$

\noindent
where $\Delta (x-q) = (x^1 - q^1, x^2 - q^2, x^3 - q^3)$ and the partials
of $F_u$ are evaluated at $\Delta q$.
Using the smoothness of $F$,
it follows that the remainder term
$R(u, q,  \Delta (x-q))$ may be bounded on some compact set
in the domain space of six variables
using the size of the second
partial derivatives  of $F$ and the
magnitude of $|\Delta (x-q)|^2$. Of course the bound will depend on
the mixed second partials of $F$ involving the $u^0, u^2, u^3$
variables as well as the $x^1, x^2, x^3$
variables, compare [9, p. 252].
In particular, one may obtain an inequality of the
form

\bigskip
\noindent
(3.1)   \qquad \qquad \qquad $| R(u, q,  \Delta (x - q)) | < M | \Delta
(x-q) |^2$
\bigskip

\noindent
where $M$ is a constant which holds for $u$ near $r$, $q$ near $p$,
and all sufficiently small $|\Delta (x-q)|$.
We will use this
bound on the size of the remainder term in the proof of Theorem 3.5 below.
We remark in passing that points of any sufficiently small neighborhood
$W(p)$ of $p = (0, 0, 0, 0)$ will lie on many of the null surfaces $S(u^0,
u^2, u^3)$.
In fact, the null
cone from each point of a sufficiently small $W(p)$ will
intersect the coordinate plane
$x^1 = t_1$ in a two dimensional surface and thus through
a point of $W(p)$ there
will be a two parameter family of null surfaces in the collection
$S(u^0, u^1, u^2)$.

The next result answers a question raised by
Chru\'sciel and Galloway  \cite{cag}.    They proved that  Cauchy horizons
fail to be differentiable
at endpoints with more than one null generator and that Cauchy horizons are
differentiable at  points which are not endpoints \cite{cag}.
They mentioned that these results
left open the question of the differentiability
of Cauchy horizons at endpoints
where there is only null generator (i.e., endpoints of multiplicity one
in our terminology). As Chru\'sciel and Galloway noted, it is of
interest to resolve this remaining differentiability question.
In Theorem 3.5,
we obtain a positive answer to  this
question.  We show that a Cauchy horizon is always differentiable at
points of multiplicity one.

\bigskip
\bigskip

\noindent {\bf Theorem 3.5.}
{\em A Cauchy horizon is differentiable at all points of multiplicity one.
In particular, a Cauchy horizon is differentiable at an endpoint where
only one null generator leaves the horizon.}

\smallskip

\noindent {\em Proof.}
Let $p$ be a point of multiplicity one. We will use the above described
coordinates $x^0, x^1, x^2, x^3$, neighborhoods $U(r)$, $N(V_0)$, $W(p)$,
null surfaces $S(u^0, u^2, u^3)$, functions $F_u(x^1, x^2, x^3)$ and
Inequality (3.1).

Assume that the Cauchy horizon $H^+(S)$ is given
near $p$ by $x^0 = f_H(\Delta x)$.
We will show that $f_H$ must be differentiable at the origin and have all
of its first order partials equal to zero at the origin.
This will prove that
$H^+(S)$ is differentiable at the origin and that it has
its tangent plane  at the origin  tangent to
the coordinate plane $x^0 = 0$.  To show the desired properties of
$f_H$ are true, it is
sufficient to show that $ f_H(\Delta x) / |\Delta x|$ converges to zero
as $| \Delta x |$ converges to zero.  Hence, by way of contradiction, we
assume that there is a sequence of
points $\{ q_k = (q_k{}^0, q_k{}^1, q_k{}^2, q_k{}^3) \}$
on $H^+(S)$ and converging to $p$
with

$$
| f_H(\Delta q_k) | / |\Delta q_k| > c > 0
$$

\smallskip

\noindent
for all $k$.   Here

$$
\Delta q_k = (q_k{}^1, q_k{}^2, q_k{}^3)  \qquad  {\rm and} \qquad
| \Delta q_k | = \sqrt{ (q_k{}^1)^2 + (q_k{}^2)^2  + (q_k{}^3)^2 }.
$$

\smallskip

\noindent
The surface
$S(0, 0, 0)$ is tangent at the origin to the
coordinate plane $x^0 = 0$ and represents
(part of) the null cone from the point $r = (0, t_1, 0, 0)$.
Note that points in the
chronological future of $r$ must lie above the Cauchy horizon and that
$\partial / \partial x^0$ is future pointing timelike at the origin.  It follows
that points of the horizon lie on or below the surface $S(0, 0, 0)$ and
hence

$$
q_k{}^0 = f_H(\Delta q_k) \leq F(0,0,0, q_k{}^1, q_k{}^2, q_k{}^3)
$$

\smallskip
\noindent
for all $k$.  Expanding
$F(0,0,0, x^1, x^2, x^3)$ about the origin in the $x$ variables, one has
$F(0, 0, 0, q_k{}^1, q_k{}^2, q_k{}^3) = 0 + 0 + R( \Delta q_k)$
where
$R( \Delta q_k) / | \Delta q_k |$ converges to zero as $k$ increases.
Using $q_k{}^0 = f_H(\Delta q_k) \leq F(0, 0, 0, \Delta q_k)$ and
$| f_H(\Delta q_k)| / | \Delta q_k |  > c$, one finds that
for all large $k$ one must have $f_H( \Delta q_k) < 0$.
Also, one obtains
the inequality $f_H(\Delta q_k) < - c |\Delta q_k |$ for all large $k$.
For each $k$, choose a null generator $\gamma_k$ containing $q_k$.
Lemma 3.1 guarantees that
the sequence $\{ \gamma_k \}$ converges to
the unique null generator $\gamma$ ( $ = x^1$ axis)
containing $p$.  Thus, for sufficiently
large $k$, each $\gamma_k$ contains points
close to $r$ and has  tangent directions for these
points  close to
that of $\gamma$ at $r$ (i.e., close to the direction of the $x^1$ axis).
It follows that for each large $k$ there will be a
well defined point $u_k$ on $\gamma_k$ with
$u_k = (u_k{}^0, t_1, u_k{}^2, u_k{}^3) \in U(r)$ and
the sequence $u_k$ will converge to $r$.
Furthermore, we may parametrize $\gamma_k$
such that $\gamma_k(0) = u_k$ and $\gamma_k(1) = q_k$.  Then
$V_k = \gamma_k{}^{\prime}(0)$ converges to $V_0$.  The null
surface $S(u_k{}^0, u_k{}^2, u_k{}^3)$ contains $q_k$ and is given by
$x^0 = F_k(\Delta x)$ where
$F_k(\Delta x) = F(u_k{}^0, u_k{}^2, u_k{}^3, x^1, x^2, x^3)$.  We will
show that for large $k$ one has
$F_k(0, 0, 0) < 0$ and thus that the surface
$S(u_k{}^0, u_k{}^2, u_k{}^3)$ cuts the $x^0$ axis below the origin.
To this end, expand $F_k$ about the point $q_k$ in the $x$
variables to obtain

$$
F_k(\Delta x) = q_k{}^0 + \sum a_{ki}(x^i - q_k{}^i) +
                 R_k(\Delta (x - q_k))
$$

\noindent
where $\Delta (x - q_k) = (x^1 - q_k{}^1, x^2 - q_k{}^2, x^3 - q_k{}^3)$,
$R_k(\Delta (x - q_k)) =  R(u_k, q_k, \Delta (x - q_k)) =
R(u_k{}^0, u_k{}^2, u_k{}^3, q^1, q^2, q^3,  \Delta (x - q_k))$
and
$a_{ki}$ represents the partial of $F_k$ with respect to $x^i$ evaluated
at the point $q_k$ for $i = 1, 2, 3$.  Using $x^1 = x^2 = x^3 = 0$ and
$q_k{}^0 = f_H(\Delta q_k) < - c |\Delta q_k|$, one obtains the following
inequality for large $k$

$$
F_k(0, 0, 0) < -c |\Delta q_k| + \sum a_{ki}(0 - q_k{}^i) +
                R_k(- q_k{}^1, - q_k{}^2, - q_k{}^3)
$$

\noindent
Recall that as $k$ increases, the
tangent direction to $\gamma_k$ converges to the
tangent direction to $\gamma$.  Thus, the
null tangent plane to $x^0 = F_k(\Delta x)$
at $q_k$ must converge to the null plane tangent to
$x^0 = 0$ at the origin.  It follows that the coefficients
$a_{ki}$ converge to zero for $i = 1, 2, 3$ and thus for large $k$ one has
an inequality of the form

$$
|\sum a_{ki}(0 - q_k{}^i)|  <   (c/3) |\Delta q_k|.
$$

\noindent
Also, using  Inequality (3.1) one has
$|R_k(- q_k{}^1, - q_k{}^2, - q_k{}^3)|  < M |\Delta q_k |^2$
which implies that for large $k$

$$
|R_k(- q_k{}^1, - q_k{}^2, - q_k{}^3)| < (c/3) |\Delta q_k |.
$$

\noindent
Consequently, we find

$$
F_k(0,0,0) < - c |\Delta q_k| + (c/3)|\Delta q_k| + (c/3)|\Delta q_k|
$$

\noindent
which yields $F_k(0,0,0) < 0$ for large $k$ since $| \Delta q_k | \neq 0$.
Using the fact that the $x^0$ axis is
timelike future directed, one finds that
$F_k(0, 0, 0)$ is in the
chronological past of $p = (0, 0, 0, 0)$ for large $k$.
Using the fact that $u_k$  is in the
causal past of $F_k(0, 0, 0)$, one finds that
$u_k$ must be in the chronological past of $p = (0, 0, 0, 0)$
for large $k$.  Since $p$ and $u_k$ must lie on $H^+(S)$, this
contradicts the achronality of the horizon.  We conclude
that $f_H$  is differentiable at the origin
and  all of its partials
with respect to $x^i$ are zero at the origin.
It follows that $H^+(S)$ is differentiable
at the origin, as desired.  $\Box$

\bigskip
\bigskip

Recall that the {\em crease set} is
the set of points in the
horizon having multiplicity at least two and that the horizon
fails to be differentiable at any such point.
Clearly, Theorem 3.5 is equivalent to the statement
that {\em the crease set consists
precisely of those points on the Cauchy horizon
where the horizon fails to
be differentiable}.

Combining Propositions 3.3  and 3.4 with  Theorem 3.5,  one
obtains the following result.

\bigskip
\bigskip

\noindent {\bf Proposition 3.6.}
{\em Let $W$ be an open subset of the Cauchy horizon $H^+(S)$.
Then the following are equivalent.

1.  $H^+(S)$ is differentiable on $W$.

2.  $H^+(S)$ is of class $C^r$ on $W$ for some $r \geq 1$.

3.  $H^+(S)$ has no endpoints on $W$.

4.  All points of $W$ have multiplicity one.}

\bigskip
\bigskip

Note that the four parts of Proposition 3.6 are logically
equivalent for an {\em open} set $W$, but that, in general, they
are not necessarily equivalent for sets which fail to be open.
Using the equivalence of parts (1) and (3) of
Proposition 3.6, it now follows that near each
endpoint of multiplicity one there must
be points where the horizon fails to be differentiable.
Hence, each neighborhood of an endpoint of multiplicity
one must contain endpoints of higher multiplicity. This yields
the following corollary.

\bigskip
\bigskip

\noindent {\bf Corollary 3.7.}
If $p$ is an endpoint of multiplicity one on a Cauchy horizon $H^+(S)$,
then each neighborhood $W(p)$ of $p$ on $H^+(S)$ contains points where
the horizon fails to be differentiable.  Hence, the set of endpoints of
multiplicity one is in the closure of the crease set.

\bigskip

\bigskip

We know
that the nondifferentiable set (i.e., the crease set) has (3-dimensional)
measure zero since the horizon satisfies a Lipschitz condition.  On the
other hand, since a Cauchy horizon is differentiable at an endpoint of
multiplicity one, the fact that $H^+(S)$ satisfies a Lipschitz condition
does not give us any direct information on the measure of the set of
endpoints of multiplicity one. Based on known examples it seems likely that
this set should be a relatively small set.
Thus, we conjecture that the set of endpoints of multiplicity one
is of measure zero for all Cauchy horizons.  Clearly, this is
equivalent to the following  endpoint conjecture.

\bigskip

\bigskip

\noindent {\bf Conjecture.}  The set of all endpoints
of a Cauchy horizon must have measure zero.

\bigskip

\bigskip

An affirmative or negative answer to the above
conjecture would certainly  help in the
understanding of Cauchy horizons and in future
studies of these horizons.

\bigskip
\bigskip

\bigskip

\centerline {\bf 4.  Compact Horizons}

\bigskip
\bigskip

 In the previous sections we have not made any curvature assumptions. In this
section we will assume $Ric (V, V) \geq 0$ for all null
$V$ which is the {\em null convergence condition}.
We will also assume that we have a Cauchy horizon
$H^{+}(S)$ which is compact and of class $C^2$ on an open subset $G$ such
that the
complement of $G$ is of 3-dimensional measure zero.

\bigskip
\bigskip

\noindent {\bf Theorem 4.1.}    {\em Let $(M, g)$ be a spacetime and assume that
$(M, g)$
satisfies the null convergence condition (i.e., $Ric (V, V) \geq 0$ for all null
$V$).  Let $S$ be a partial Cauchy surface.   Assume that the future
horizon $H^{+}(S)$ is compact and contains an open set $G$ where it is $C^2$
and is such that the complement of $G$ in $H^{+}(S)$ has 3-dimensional measure
zero.  Then, $H^{+}(S)$ has no endpoints and hence is
differentiable of (at least) class $C^1$ for all points.}

\smallskip

\noindent {\em Proof.}   The compactness of $H^+(S)$ yields the past
completeness of
all null generators [12, p. 295].  We use
the notation of \cite{haa} to obtain a map
$u_t: G \rightarrow G$, see [11, p. 606] and equation

\bigskip

\noindent (4.1)  \qquad \qquad  \qquad \qquad $d/dt \int_{ u_t(G)}  dA = 2
\int_{ u_t(G)} \rho dA$

\bigskip

\noindent
Notice that since $H^{+}(S)$ is compact we find that $\int_{ u_t(G)}
dA$  is finite.  However,  the derivative of
$ \int_{ u_t(G)}  dA$
cannot be positive since the set $G$ is mapped into itself.  Thus, the
left hand side of Equation 4.1 is nonpositive.  On the other hand, the
right hand side of Equation 4.1 must be nonnegative since $\rho \geq 0$.  Assume
now that $H^{+}(S)$ has an endpoint $p$ of a null generator $\gamma$.
Even if $\gamma$ does
not lie in $G$, the horizon will be differentiable on the part of
$\gamma$ in the past of $p$.  Choose some
$q$ on $\gamma$ in the past of $p$ and some $u$ on $\gamma$ in the future
of $p$.
Then for
some small neighborhood $W(q)$ of $q$ on $H^{+}(S)$, all null generators though
points $r$ of $W(q)$ will have directions close to the direction of $\gamma$ at
$q$.   Recall that given a compact domain set in the $t-$axis, geodesics
with close initial conditions remain close on the compact domain set.
Thus, by choosing $W(q)$ sufficiently small we may get all null
generators though points of $W(q)$ come arbitrarily close to $u$.  Thus
for sufficiently small $W(q)$ all null generators intersecting $W(q)$ must
leave $H^{+}(S)$ in the future.  Since $W(q)$ is open in $H^{+}(S)$ it must
have a
nontrivial intersection with the set $G$.  Thus, $u_t(G)$ cannot be all of $G$
for some positive values of $t$ and this yields a negative for some values of
$t$ on the left hand
side of Equation 4.1, in contradiction. Thus, $H^+(S)$ has no endpoints and by
Proposition 3.6 must be (at least) of class $C^1$ at all points.  $\Box$

\bigskip

\bigskip

 \noindent {\bf Acknowledgment:} The authors wish to thank P. T. Chru\'sciel and
                 G. Galloway for helpful discussions. A. K. wishes to thank
                 Albert Einstein Institut for hospitality where part of this 
                 work was done and acknowledge the support of the Polish Science
                 Committee through grant KNB 2 P301 050 07.
                 Furthermore, J. K. B. wishes to
                 thank the Erwin Schr\"odinger Institute of
                 Mathematical Physics for partial support.

\bigskip

\bigskip


\end{document}